\documentclass[cits]{PoS}
\usepackage{amsmath}

\def \be {\mathbf{E}}

\def \br {\mathbf{r}}

\def \cf {C_F}
\def \nc {N_c}

\def \bp {\mathbf{p}}

\newcommand{\Tint}[1]{{\hbox{$\sum$}\!\!\!\!\!\!\!\int\,}_{\!\!\!\!\raise-0.9ex\hbox{$\scriptstyle{#1}$}}}

\def\siml{{\ \lower-1.2pt\vbox{\hbox{\rlap{$<$}\lower6pt\vbox{\hbox{$\sim$}}}}\ }}
\def\simg{{\ \lower-1.2pt\vbox{\hbox{\rlap{$>$}\lower6pt\vbox{\hbox{$\sim$}}}}\ }}

\def \be {\mathbf{E}}

\def \br {\mathbf{r}}

\def \bp {\mathbf{p}}

\def \als {\alpha_{\mathrm{s}}}

\def \m2   {\mu^{2 \epsilon}}

\def\siml{{\ \lower-1.2pt\vbox{\hbox{\rlap{$<$}\lower6pt\vbox{\hbox{$\sim$}}}}\ }}
\def\simg{{\ \lower-1.2pt\vbox{\hbox{\rlap{$>$}\lower6pt\vbox{\hbox{$\sim$}}}}\ }}
\def\lqcd{\Lambda_\mathrm{QCD}}

\title{Review of the EFT treatment of quarkonium at finite temperature}

\ShortTitle{Review of the EFT treatment of quarkonium at finite temperature}

\author{\speaker{Jacopo Ghiglieri}\\
       McGill University, Department of Physics,\\
	3600 rue University, Montreal QC H3A 2T8, Canada\\
       E-mail: \email{jacopo.ghiglieri@physics.mcgill.ca}}

\abstract{Heavy quarkonium is one of the most investigated probes of the medium produced in heavy-ion collisions. In the past few years progress has been made in the description of its in-medium dynamics from QCD. Non-relativistic EFTs in particular allow one to define and compute rigorously the potential governing the real-time evolution of the bound state from QCD, showing that it is composed of a real and an imaginary part. The latter encodes the effect of scattering with the light constituents of the plasma and is oftentimes responsible for the dissociation of the bound state. The EFT approach and its main results, as well as comparisons with recent lattice data and phenomenological applications, are briefly reviewed here.}

\FullConference{Xth Quark Confinement and the Hadron Spectrum\\
		 8-12 October 2012\\
		 TUM Campus Garching, Munich, Germany}

\begin{document}

\section{Introduction}
The phase diagram of QCD in the region of vanishing chemical potential is actively 
explored in heavy ion collision experiments at RHIC and LHC. Lattice calculations 
(see \cite{Petreczky:2012rq} for a recent review) predict a crossover transition 
to a larger number of degrees of freedom, typical of a deconfined medium (the Quark-Gluon
Plasma), for 
temperatures $T$ ranging from 150 to 200 MeV.  On the experimental side, the
characterization of the properties of the medium relies either on its bulk properties,
which find an effective description in hydrodynamics, or  in 
\emph{hard probes}, i.e. energetic particles not in equilibrium
with the medium. 

Heavy quarkonium has been one of the most actively investigated hard probes for the
past 26 years. In 1986 Matsui and Satz \cite{Matsui:1986dk} hypothesized that colour
screening in a deconfined medium would have dissociated the $J/\psi$, resulting
in a suppressed yield in the easily accessible dilepton decay channel and yielding
a striking QGP signature. Such a suppressed yield has been indeed observed in 
heavy ion collision experiments at SPS, RHIC and LHC (see \cite{Brambilla:2010cs}
for a review), the current understanding being that it cannot be explained by
cold nuclear matter effects alone, i.e. those caused by a confined nuclear medium.
Furthermore the LHC results have opened up the frontier of the cleaner bottomonium 
probe with the availability of quality data on the $\Upsilon$ resonances (see 
\cite{CMSupsi} for the latest CMS results).

On the theory side, a great deal of the studies of the in-medium dynamics of 
the $Q\overline{Q}$ bound states has been carried out with phenomenological
 potential models, first introduced in \cite{Karsch:1987pv}, where all medium 
effects are encoded in a $T$-dependent potential plugged in a Schr\"odinger 
equation.  We refer to \cite{Rapp:2008tf,Kluberg:2009wc,Bazavov:2009us} for 
recent reviews. The derivation of such models from QCD was however not established. 
Moreover, lattice calculations of free energies and other quantities 
\cite{McLerran:1981pb,Nadkarni:1986as} obtained from correlation functions 
of Polyakov loops are often taken as input for the $T$-dependent potential. 
Although these quantities have been thought to be related to the colour-singlet 
and colour-octet heavy quark potentials at finite temperature 
\cite{Nadkarni:1986cz,Nadkarni:1986as}, a precise relation  was still 
lacking in the literature, as pointed out in \cite{Philipsen:2008qx}.

In the past few years, a considerable effort has been devoted to the derivation
of a proper potential from QCD. In \cite{Laine:2006ns} (see also \cite{Beraudo:2007ky}) it was shown that the perturbative
real-time potential obtained by analytical continuation of a rectangular Wilson
loop shows a screened real part and a large imaginary part (for $T\gg 1/r\sim m_D$,\
 $m_D$ being the Debye mass, the imaginary part is much larger than
the real part) which is due to the scattering of the virtual potential gluons
with the light particles (quarks and gluons) of the medium, in what is called 
\emph{Landau damping}. Further phenomenological studies have shown 
the numerical importance of this imaginary part \cite{Laine:2007gj,Burnier:2007qm,Petreczky:2010tk}
when solving a Schr\"odinger equation. A generalization to anisotropic plasmas has been
performed in \cite{Burnier:2009yu,Dumitru:2009fy,Philipsen:2009wg} and used in \cite{Strickland:2011mw,Strickland:2011aa} (coupled with a hydrodynamical evolution)
 to obtain predictions on the suppression rates.

The first-principle derivation of potentials from QCD was further improved with 
the development of an Effective Field Theory (EFT) framework \cite{Brambilla:2008cx,Escobedo:2008sy,Brambilla:2010vq,Brambilla:2010xn,Escobedo:2010tu,Escobedo:2011ie,Brambilla:2011mk,arXiv:1109.5826,Berwein:2012mw},
 which will be at the center of this 
contribution. This approach, started in \cite{Brambilla:2008cx} (see also
\cite{Escobedo:2008sy} for the abelian analog), relies on the well-established
framework of $T=0$ Non-Relativistic (NR) EFTs of QCD and extends it to
finite temperatures, giving to the potential a modern, rigorous definition
as a matching coefficient of the EFT that arises when the scale $mv$ is integrated out,
where $m$ is the heavy quark ($Q$) mass and $v\ll 1$ is the relative velocity of the constituents of the NR bound state.
At zero temperature this EFT is called Potential Non-Relativistic QCD (pNRQCD) \cite{Pineda:1997bj,Brambilla:1999xf}.

This contribution is structured as follows: in Sec.~\ref{sec_eft} we will
illustrate the basic principles of the EFT framework. In sections~\ref{sec_screen}
and \ref{sec_pert} we will apply those principles to the real-time dynamics
of bound states in two interesting temperature regimes, $T>mv$ and $T<mv$ respectively.
In Sec.~\ref{sec_cross} we will compare the thermal widths obtained within the
EFT framework with other approaches in the literature and show how dissociation
cross sections can be extracted from the EFT calculations. In Sec.~\ref{sec_thermo}
we will deal with Euclidean quantities and their related thermodynamical free energies,
showing how they relate to the $Q\overline{Q}$ potentials. In Sec.~\ref{sec_aspect}
we will briefly review some interesting recent developments and finally in
Sec.~\ref{sec_concl} we will draw our conclusions.

\section{EFTs}
\label{sec_eft}
In a scenario where a physical system is described by two scales $\Lambda\gg m$, 
a low-energy EFT describing the physics at the scale $m$ is constructed by 
\emph{integrating out $\Lambda$}, i.e. integrating out modes of energy/momentum
 above a given cut-off $\mu$, $\Lambda\gg\mu\gg m$, and imposing that the EFT and the theory
 it is derived from are equivalent below the cut-off. If the system is characterized
by more than two scales and if these scales are hierarchically ordered, the procedure
can be iterated, creating a \emph{tower} or \emph{hierarchy of EFTs}. If two scales are not hierarchically
separated, they should then be integrated out in the same step.

This line of action is at the base of the above-mentioned $T=0$
NR EFT framework. As we remarked, the expansion parameter there is $v\ll1$, 
giving rise to the hierarchy of scales $m\gg mv\gg mv^2$, 
where $m$ is the heavy quark mass, $mv$ is the typical relative momentum
or inverse radius and $mv^2$ the kinetic/binding energy. The framework 
is constructed by integrating out first the mass scale, yielding Non-Relativistic QCD 
(NRQCD) \cite{Caswell:1985ui,Bodwin:1994jh}. Its Lagrangian is organized 
as an expansion in the inverse mass. The second step is the integration
of the scale $mv$, which leads to pNRQCD \cite{Pineda:1997bj,Brambilla:1999xf}. The relative hierarchical position
of $\lqcd$ and $mv$ establishes whether this integration is to be performed
perturbatively or non-perturbatively. The former case leads to weakly-coupled
pNRQCD, which we illustrate briefly because the EFTs
developed at finite temperature, which will be introduced later, will share
its basic form.

The degrees of freedom in the $Q\overline{Q}$ sector are conveniently
written as a colour-singlet and a colour-octet $Q\overline{Q}$ bilinear
field, which can interact through ultrasoft ($E,p\sim mv^2$ gluons).
This cutoff is effectively expressed through a multipole expansion. In
detail, the Lagrangian reads \cite{Pineda:1997bj,Brambilla:1999xf}
\begin{eqnarray}
 %\hspace{-1.5cm}
{\cal L}_{\textrm{pNRQCD}} &=& - \frac{1}{4} F^a_{\mu \nu} F^{a\,\mu \nu} + \sum_{i=1}^{n_f}\bar{q}_i\,iD\!\!\!\!/\,q_i
%\nonumber\\ 
%&&\hspace{0.5cm}
+ \int d^3r \; {\rm Tr} \,  
\Bigl\{ {\rm S}^\dagger \left[ i\partial_0 - h_s \right] {\rm S} 
+ {\rm O}^\dagger \left[ iD_0 -h_o \right] {\rm O} 
\nonumber \\
&& \hspace{1.8cm}
+ V_A\, \left( {\rm O}^\dagger \br \cdot g\be \,{\rm S} + \textrm{H.c.} \right)
+ \frac{V_B}{2} {\rm O}^\dagger \left\{ \br\cdot g\be \,, {\rm O}\right\} 
+ \dots\,\Bigr\} .
\label{pNRQCD}	
\end{eqnarray}
The fields $\mathrm{S}=S\,\mathbf{1}_c/\sqrt{N_c}$ and $\mathrm{O}=O^a\,T^a/\sqrt{T_F}$ are the $Q\overline{Q}$ 
colour-singlet and colour-octet fields respectively, $q_i$ are the light quarks, in $n_f$ flavours,
$T_F=1/2$,
$\be$ is the chromoelectric field,  $iD_0 \mathrm{O} =i\partial_0 \mathrm{O} - gA_0 \mathrm{O} + \mathrm{O} gA_0$ 
and H.c. stands for Hermitian conjugate. The trace is over colour and spin indices. 
%Gluon fields depend only on the centre-of-mass coordinate and on time;
%this is a consequence of the aforementioned multipole expansion in the quark-antiquark relative
%distance $r$. 
The dots in the last line stand for higher-order terms in $r$ and $1/m$.

The dependence on the scales $m$ and $mv\sim m\als$ is encoded in the Wilson coefficients; $V_A$
and $V_B$ are at leading order $V_A=V_B=1$, whereas the singlet and octet Hamiltonians have the form 
($\bp\equiv-i\nabla_\br$) 
\begin{equation}
h_s =\frac{\bp^2}{m}-\cf\frac{\als}{r}+\ldots,\qquad 
h_o =\frac{\bp^2}{m}+\frac{1}{2\nc}\frac{\als}{r}+\ldots.
\label{leadinghams}
\end{equation}
The dots in these equations stand for higher orders in the $1/m$ and $\als$
expansions, such as spin- and momentum-dependent terms or radiative corrections
to the Coulomb potentials. $\nc=3$ is the number of colours and $\cf=4/3$ in QCD. In the power-counting
of the EFT the explicitly-shown terms are of the same size and give rise to 
a Coulombic bound state. At the zeroth order in the multipole expansion the 
equation of motion for the singlet field arising from Eq.~\eqref{pNRQCD} is 
then a simple Schr\"odinger equation, resulting in a spectrum
 made by the (QCD) Bohr levels $E_n = -4m\als^2/(9n^2)$, 
whereas the octet potential is repulsive and does not support bound states but 
a continuum of scattering states.  Higher orders in the multipole expansion,
such as the chromoelectric dipole couplings on the second line of Eq.~\eqref{pNRQCD},
are responsibile for retardation effects such as those yielding the
Lamb shift in hydrogen spectroscopy.

At finite temperature one also encounters the scales that characterize
the thermal medium: its temperature $T$, the electric (Debye) screening
mass $m_D\sim gT$ and a magnetic mass of order $g^2T$.\footnote{We do
not distinguish between $T$ and $\pi T$ or multiples thereof in the text.
For what concerns the magnetic mass, we never reach an accuracy
where its contribution becomes relevant, so we need not worry about
its non-perturbative treatment.} In the weak-coupling scenario
that has so far been investigated in the EFT approach, these scales
develop a hierarchy. In order to construct a finite temperature
EFT framework analogous to the one just illustrated, one needs then
to put these scales together with the NR ones in a global hierarchy.
Many such hierarchies are possible and phenomenologically sensible. 
The latter requirement translates into $T\ll m$. 

Once a hierarchy is fixed, one proceeds to integrate out scales
along the procedure sketched above. Since $m\gg T$, the first step
is always the integration of the mass, leading again to NRQCD. Since 
an EFT and its matching coefficients are independent of the hierarchy
of scales below the cutoff, its Lagrangian is identical to the zero-temperature
case.

After this first step, the subsequent ones differ in the different
cases, which in the following we will group into two regions: the 
one where $T\gg mv$ and the one where $T\ll mv$, which can be of relevance
for the ground state of bottomonium. The case $T\sim mv$ has been 
dealt with in QED in \cite{Escobedo:2010tu}; its QCD investigation
is underway in \cite{quasifree}. When integrating out the thermal
scales, existing finite-T EFTs of QCD such as the Hard Thermal Loop (HTL) 
effective theory \cite{Braaten:1991gm} are employed.

In all cases, once the scale $mv$ is integrated out, the resulting 
Lagrangian is similar to the one of pNRQCD, with  colour-singlet 
and  colour-octet fields and a Schr\"odinger equation picture appearing as the
zeroth-order equation of motion. The potential
is then rigorously defined as a matching coefficient; all scales
above $mv^2$ will contribute to it.

\begin{figure}[ht]
	\begin{center}
		\includegraphics[width=10cm]{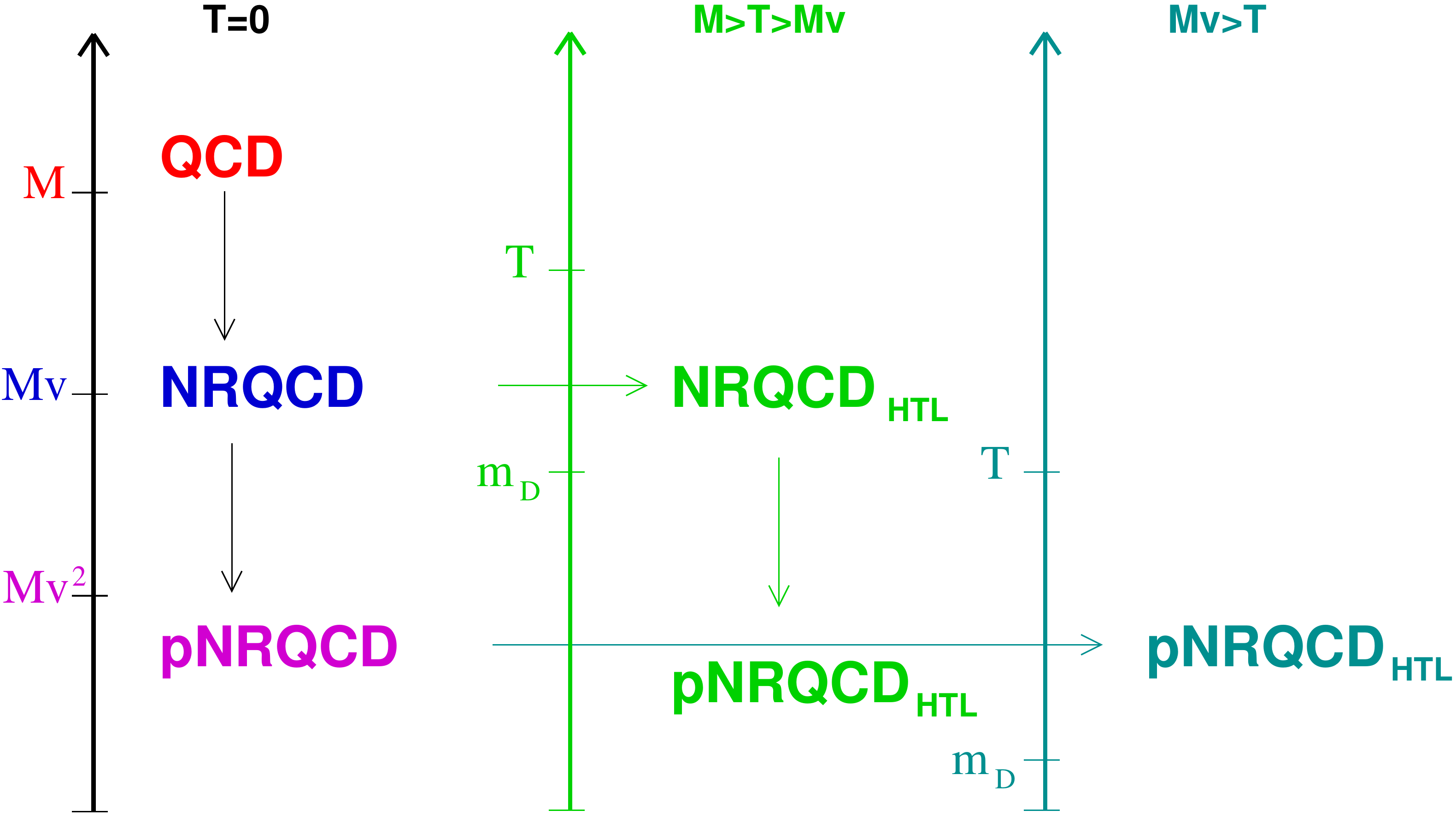}
	\end{center}
	\caption{Possible hierarchies and the EFTs they generate.
	Figure taken from \cite{Vairo:2011az}.}
	\label{fig_antonio}
\end{figure}

The different possibilities for the hierarchies and the EFTs
that arise out of them are pictured in Fig.~\ref{fig_antonio}.
We conclude this Section by noting that the NR EFT framework inherits all the
strong points of EFTs. In particular, it has a well-defined power counting,
which allows for easy estimations of the size of possible contributions
to interesting observables, and it is systematically improvable, i.e. new
operators and matching coefficient can be added to the Lagrangian and computed
in order to reach the desired accuracy in the calculation of the above-mentioned
observables.
\section{The $m\gg T\gg mv$ region}
\label{sec_screen}
In this region the temperature scale is to be integrated first, obtaining 
an EFT of NRQCD called NRQCD${}_\mathrm{HTL}$ \cite{Vairo:2009ih,Ghiglieri:2012rp},
where the gauge and light-quark sector is described by the HTL Lagrangian.
The second step depends on the existence of a hierarchical separation 
between $m_D$ and $mv$. If $m_D\sim mv$ the two scales are to be integrated out
at the same time. If, on the other hand, $mv\gg m_D$, one needs two separate steps. 
For reasons that will become clearer later, the case $mv\ll m_D$ is not relevant.

In the first case one obtains a pNRQCD-like EFT named pNRQCD${}_\mathrm{HTL}$,
where the potential, obtained by
matching from NRQCD${}_\mathrm{HTL}$ in the real-time formalism of Thermal Field 
Theory, reproduces the earlier result of \cite{Laine:2006ns}, thus bringing it
in a consistent EFT picture \cite{Brambilla:2008cx}. This derivation
 furthermore shows the
computational advantage of the real-time formalism over the imaginary-time one
for this class of Minkowskian observables. The colour-singlet potential reads
\begin{equation}
V_s(r)\big\vert_{T\gg mv\sim m_D} =  -C_F\,\frac{\als}{r}\,e^{-m_Dr} -C_F\als m_D
+ i2C_F\,\als\, T\,\int_0^\infty dt \,\left(\frac{\sin(m_Dr\,t)}{m_Dr\,t}-1\right)\frac{t}{(t^2+1)^2}
\,,
\label{Vsrsimmd}
\end{equation}
where $m_D^2=g^2 T^2(1+n_f/6)$.
We remark that the real part scales like $\als m_D$ (recall that $mv\sim 1/r\sim m_D$). The imaginary
part is Bose-enhanced and scales like $\als T$; it is thus much larger, as we mentioned
before. Hence, one can expect
bound states to be dissociated in this regime.

If one moves to the second possibility $mv\gg m_D$, the integration procedure has to go
through an intermediate step. The real part is dominated by the $T=0$ Coulomb term, 
whereas the imaginary part at the scale $mv$ is IR divergent and 
is regulated in dimensional regularization. Once the UV-divergent contribution of the smaller scale $m_D$ is added up, the following finite result arises \cite{Brambilla:2008cx}
\begin{equation}
V_s(r)\big\vert_{T\gg mv\gg m_D} =  -C_F\,\frac{\als}{r} -\frac{C_F}{2}\als r m_D^2
+i\frac{C_F}{6}\als r^2 T m_D^2\left(\ln(rm_D)^2+2\gamma_E-\frac83\right),
\label{Vsrggmd}
\end{equation}
where $\gamma_E$ is the Euler-Mascheroni constant. As pointed out in \cite{Laine:2008cf} 
(see also \cite{Escobedo:2008sy} for QED), the real and imaginary parts of this potential
become parametrically of the same size when $T\sim m\als^{2/3}$.\footnote{The Coulombic real part
implies a Coulombic power counting, with $mv\sim1/r\sim m\als$. Logarithms of 
the coupling are not considered in this estimate} This then constitutes
a parametric estimate of the \emph{dissociation temperature}, where the
dissociation criterion popularized by potential models, i.e. disappearance 
of the binding energy, is replaced by the equality of binding energy and thermal
width. A numerical estimate has been performed in \cite{Escobedo:2010tu} for the
$\Upsilon(1S)$ and is
reported in Table~\ref{tab_miguel}.
\begin{table}[htb]
\makebox[6cm]{\phantom b}
\begin{center}
\begin{tabular}{|c|c|}
\hline
$m_c$ (MeV) & $T_d$ (MeV) \\
\hline
$\infty$ & 480 \\
5000 & 480 \\
2500 & 460 \\
1200 & 440 \\
0 & 420 \\
\hline
\end{tabular}
\end{center}
\caption{Dissociation temperature for the $\Upsilon(1S)$ for different values of the charm mass. Table and results are taken from \cite{Escobedo:2010tu}.}
\label{tab_miguel}
\end{table}
\section{The  $mv\gg T \gg mv^2$ region}
\label{sec_pert}
In this region the thermal medium acts as a perturbation to 
a Coulombic bound state, but it still modifies the potential. Furthermore,
as argued in \cite{Vairo:2010bm}, hierarchies where $mv> T> mv^2$
could be of relevance for the ground states of bottomonium, where 
$mv\sim m\als\sim 1.5$ GeV and $T$ or even $\pi T$ are smaller in
current experiments. 

In this region one starts by integrating out $mv$ from NRQCD, obtaining
standard pNRQCD. The next step, the integration of the temperature, leads
to a different version of pNRQCD${}_\mathrm{HTL}$ than the one considered 
above. Its Lagrangian has been written and some of its matching coefficients have 
been computed in \cite{Brambilla:2010vq,Brambilla:2011mk}.

Further steps depend on the hierarchy (if any) between $mv^2$ and $m_D$.
In \cite{Brambilla:2010vq} the case $mv^2\gg m_D$ has been explored in detail,
leading to a calculation of the thermal modifications to the spectrum and of the
thermal width up to order $m\als^5$ in the power-counting of the EFT. Before we 
briefly illustrate these results, we remark that other hierarchical possibilities 
have been considered in the static case in \cite{Brambilla:2008cx} and in QED in
\cite{Escobedo:2010tu}.

\begin{figure}[ht]
	\begin{center}
		\includegraphics[scale=0.6]{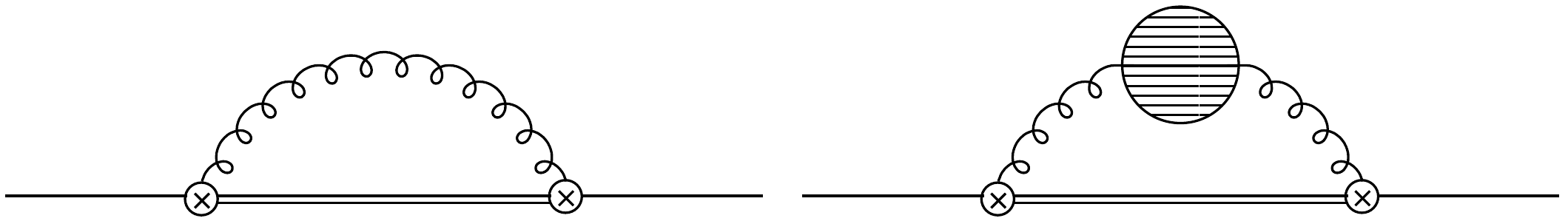}
	\end{center}
	\caption{Loop diagrams in pNRQCD and pNRQCD${}_\mathrm{HTL}$ contributing to our calculation. Single lines are colour-singlet $Q\overline{Q}$ states, double lines are colour octets, curly lines are gluons, vertices are chromoelectric dipoles and the blob is the gluon self-energy. The imaginary part of the first diagram yields the singlet-to-octet decay mechanism, whereas the second one gives the Landau damping contribution to the width.}
	\label{fig_diagrams}
\end{figure}

In order to obtain the spectrum and width to order $m\als^5$, as well as to
compute the matching coefficients of pNRQCD${}_\mathrm{HTL}$, one has to compute
loop diagrams in pNRQCD with loop momentum of order $T$ or in pNRQCD${}_\mathrm{HTL}$
with loop momenta of order $mv^2$ or $m_D$. These loop diagrams arise from the dipole
interaction vertices in Eq.~\eqref{pNRQCD}; the relevant diagrams for the
desired accuracy are shown in Fig.~\ref{fig_diagrams}.
As in the previous section, we encounter intermediate UV or IR divergences
which we regulate in dimensional regularization. The potential, which in the 
EFT framework is not an observable, shows IR divergences. The spectrum and 
the width, being instead observables, are of course finite, the divergences
cancelling in the sum of the contributions from different scales.

The shift to the binding energy induced by the thermal medium, which translates into 
a mass shift for the bound state, reads to order $m\als^5$
\begin{eqnarray}
\nonumber
\delta E_{1S}^{(\mathrm{thermal})}\big\vert_{mv\gg T\gg mv^2}&=&
\frac{34}{27}\pi\,\als^2 \,T^2 a_0
\\
\nonumber&&
+\frac{E_1\als^3}{\pi}\left[\frac{7225}{324}\left(\log\frac{4\pi^2 T^2}{E_1^2}
-2\gamma_E\right)+
\frac{128}{81}L_{1S}\right]
\\
&&
+ 3a_0^2\,\als \,T\
\left\{\frac{8}{3} \zeta(3)\,  \als \, T^2
- \left[\frac{3}{2\pi} \zeta(3)+\frac{\pi}{3}\right] \frac43 m_D^2
\right\},
\label{finalspectrum}
\end{eqnarray}
where $a_0=3/(2m\als)$ is the Bohr radius and $L_{1S}$ is the QCD equivalent
of the Bethe log in the hydrogen spectrum. We refer to \cite{Kniehl:1999ud,Kniehl:2002br}
for its definition and evaluation, which yields $L_{1S}=-81.5379$.

The thermal width reads instead
\begin{eqnarray}
\nonumber
\Gamma_{1S}^{(\mathrm{thermal})}\big\vert_{mv\gg T\gg mv^2}&=&
\frac{1156}{81}\als^3T+\frac{7225}{162}E_1\als^3
\\
\nonumber&&
-\frac{4}{3}a_0^2\als T\left[ m_D^2
\left(\ln\frac{E_1^2}{T^2}+ 2\gamma_E -3 -\log 4- 2 \frac{\zeta^\prime(2)}{\zeta(2)} \right)
+8\pi \ln 2 \; \,  \als\, T^2 \right] 
\\
&&
+\frac{32}{9}\als\, Tm_D^2\,a_0^2
\,I_{1S}\;,
\label{finalwidth}
\end{eqnarray}
where $I_{1S}=-0.49673$ is another Bethe logarithm. We refer to \cite{Brambilla:2010vq} for details.
The terms on the first line of this equation arise from the imaginary part of the diagram on
the left of Fig.~\ref{fig_diagrams} and correspond to the absorption of a thermal gluon by the 
bound state and its subsequent decay into a colour-octet state, a process we call 
\emph{singlet-to-octet decay}. The terms on the second and third lines, on the other
hand, arise from the imaginary part of the second diagram and are instead due to 
scattering of the virtual gluon with the light constituents, i.e. to Landau damping.
In the power counting of pNRQCD${}_\mathrm{HTL}$ singlet-to-octet decay
is dominant over Landau damping as long as $mv^2>m_D$, as in this case. 
We will return to these processes in the next section.

\begin{figure}[ht]
	\begin{center}
		\includegraphics[width=7.5cm]{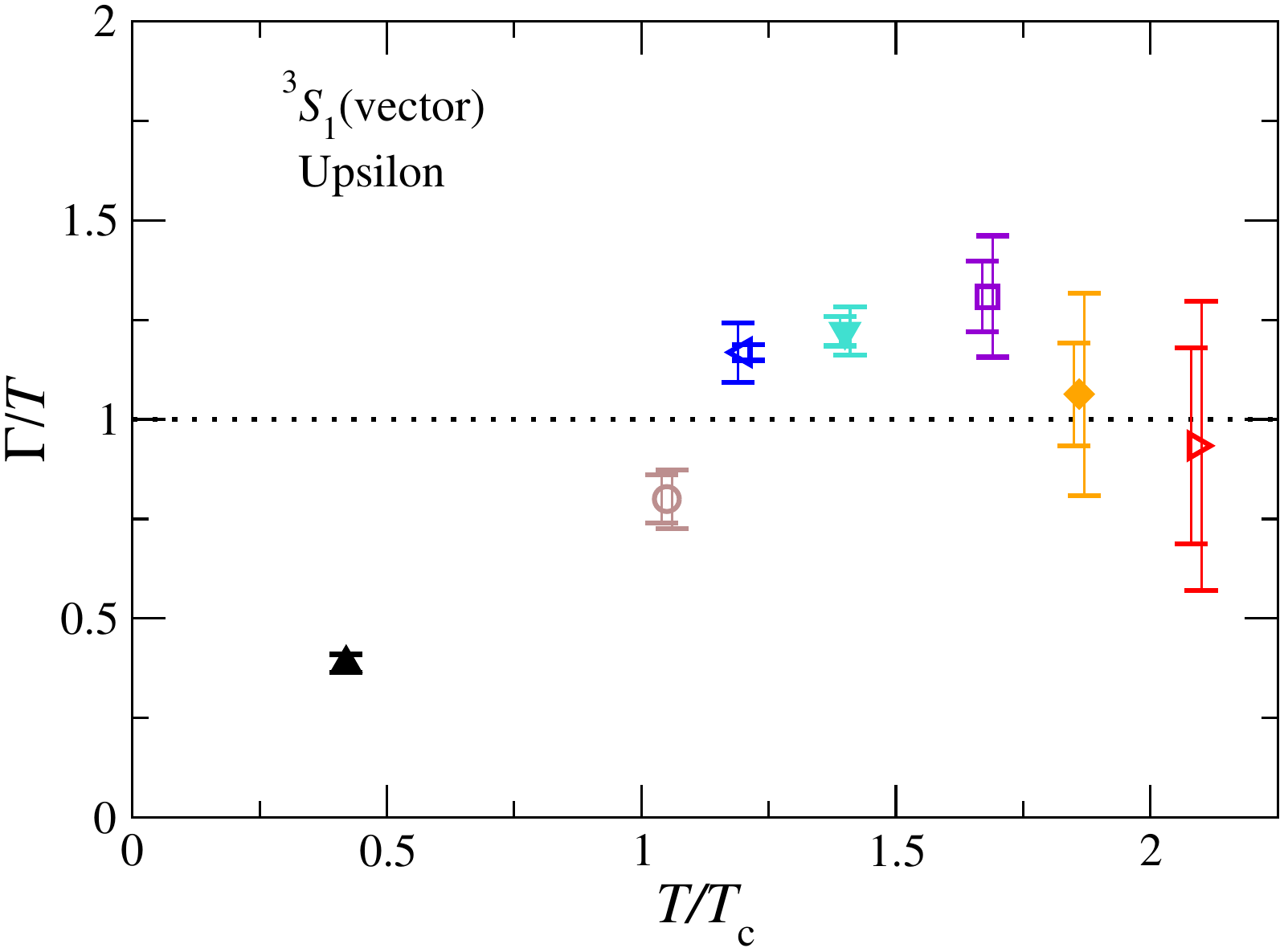}
		\includegraphics[width=7.5cm]{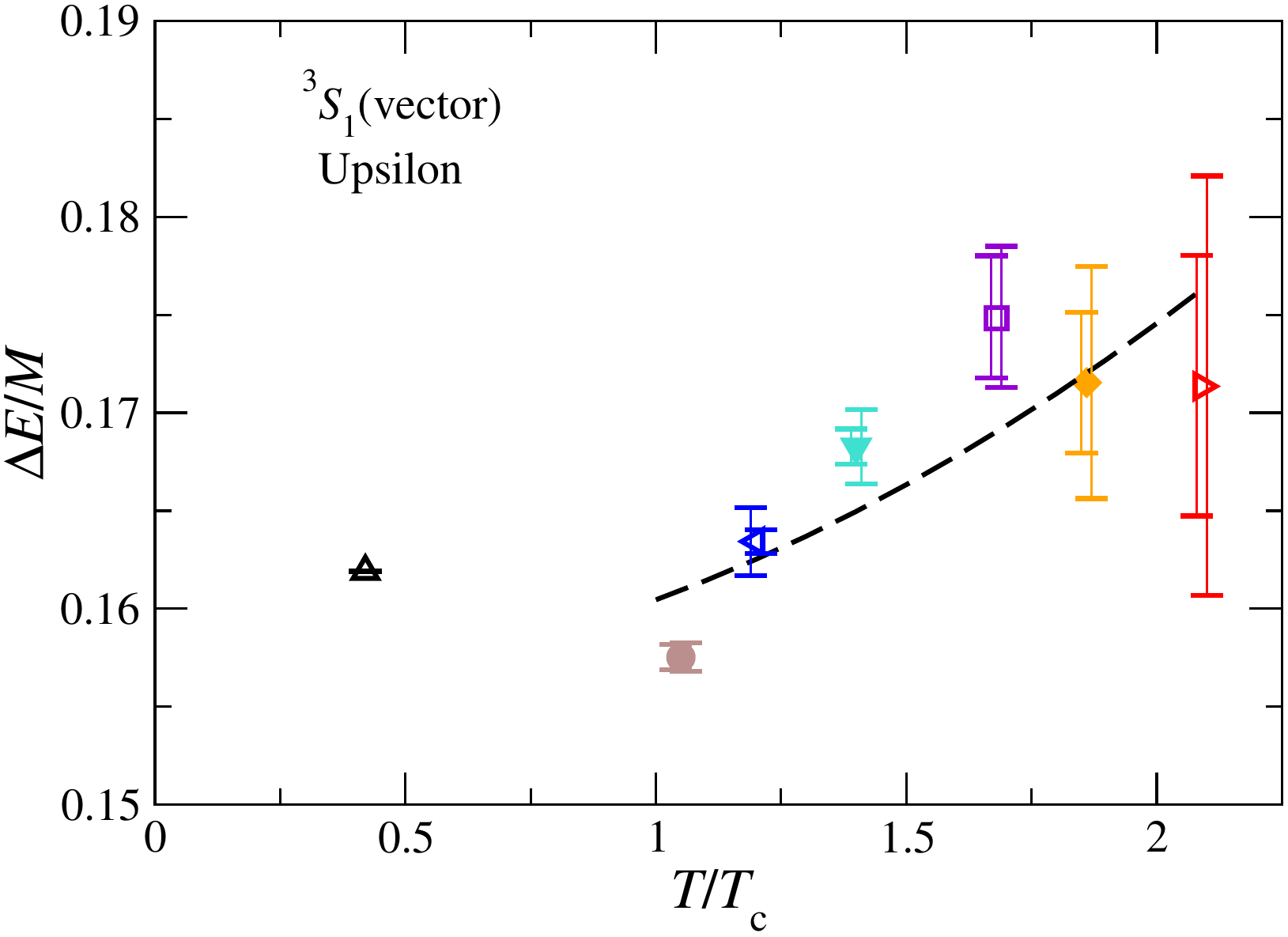}
	\end{center}
	\caption{Plots of $\Gamma$ and $\delta E$ in the vector ($\Upsilon$) channel
	from the lattice
	calculation of \cite{Aarts:2011sm}. Figures taken from that reference. The
	dashed line on the right is the leading term of our EFT calculation \cite{Brambilla:2010vq}, $\frac{34}{27}\pi\,\als^2 \,T^2 a_0$, for $m_b=5$ GeV and $\als=0.4$.}
	\label{fig_lattice}
\end{figure}

We conclude this section by reporting on the findings of \cite{Aarts:2010ek,Aarts:2011sm} (see also \cite{Ryan}). There,
the authors employed the lattice formulation of NRQCD to access $b\overline{b}$ 
bound states on the lattice at finite temperature. In \cite{Aarts:2011sm}
they used Bayesian MEM techniques to reconstruct the spectral function and 
thus extract $\delta E$ and $\Gamma$ as a function of the temperature. These
results are shown in Fig.~\ref{fig_lattice}. The authors then proceeded to
compare their results with the leading term of the EFT results, represented by
the first term on the r.h.s. of Eqs.~\eqref{finalspectrum} and \eqref{finalwidth}.
They found them to be compatible for $m_b=5$ GeV and $\als\approx0.4$.

\section{The thermal width in the EFT and in the literature}
\label{sec_cross}
In the previous sections we have identified two processes
that are responsible for the appearance of the thermal width:
these are the singlet-to-octet decay ($g+\Psi\to(Q\overline{Q})_8$, 
where $\Psi$ is the bound state and $(Q\overline{Q})_8$ the 
$Q\overline{Q}$ pair in the colour-octet state) and Landau damping
($p+\Psi\to p+(Q\overline{Q})_8$, where $p=g,q,\overline{q}$ is a 
light parton). Analogous processes had been previously considered
in the literature, going under the names of \emph{gluo-dissociation}
\cite{Kharzeev:1994pz,Xu:1995eb}
and \emph{dissociation by inelastic parton scattering} \cite{hep-ph/0103124,Grandchamp:2002wp} (see \cite{Zhao:2010nk,Emerick:2011xu,Song:2011ev} for recent applications) respectively. 
In these approaches a $T=0$ cross section for the process at hand, sometimes
complemented by momentum-independent thermal masses, was convoluted over a thermal
distribution for the incoming parton, i.e.
\begin{equation}
\Gamma=\int_{q_\mathrm{min}}\frac{d^3q}{(2\pi)^3}f_p(q)\,\sigma(q) v_\mathrm{rel}\,,
\label{eq:facw}
\end{equation}
where $f_p$ is either the Fermi--Dirac or Bose--Einstein distribution, $q$
is the incoming momentum, with kinematic minimum for dissociation $q_\mathrm{min}$,
and $v_\mathrm{rel}$ is the relative velocity. Cross sections employed in the literature
have been the Bhanot--Peskin one \cite{Bhanot:1979vb} for gluo-dissociation and \cite{CERN-TH-2574}
or more recently \cite{Song:2005yd,Park:2007zza} for  dissociation by inelastic parton 
scattering.

In \cite{arXiv:1109.5826,quasifree} (see also \cite{miguel}) we have undertaken
a comparison between the EFT results and these phenomenological approaches. For
what concerns gluo-dissociation, in \cite{arXiv:1109.5826} we have shown that our
results (such as the first line of Eq.~\eqref{finalwidth}) correspond 
to the formulation of Eq.~\eqref{eq:facw} and that the Bhanot--Peskin cross
section of \cite{Bhanot:1979vb} represents a large-$N_c$ limit of the one that
can be extracted from the EFT framework, which in turn correctly incorporates the effect
of the repulsive octet potential on the final state. An equivalent cross section has also been
derived in \cite{Brezinski:2011ju} (see also \cite{Nendzig:2012cu,Wolschin}). The 
EFT approach and its power counting clearly constrain the validity region of the 
gluo-dissociation approach to temperatures smaller than $mv$, so that the dipole approximation
employed is valid, and larger than $mv^2$, the threshold for dissociation. If the screening
mass becomes of the same order or larger than this latter scale, Hard Thermal Loop resummation
becomes necessary and the cross section is no longer temperature-independent.

\begin{figure}[ht]
	\begin{center}
		\includegraphics[width=10cm]{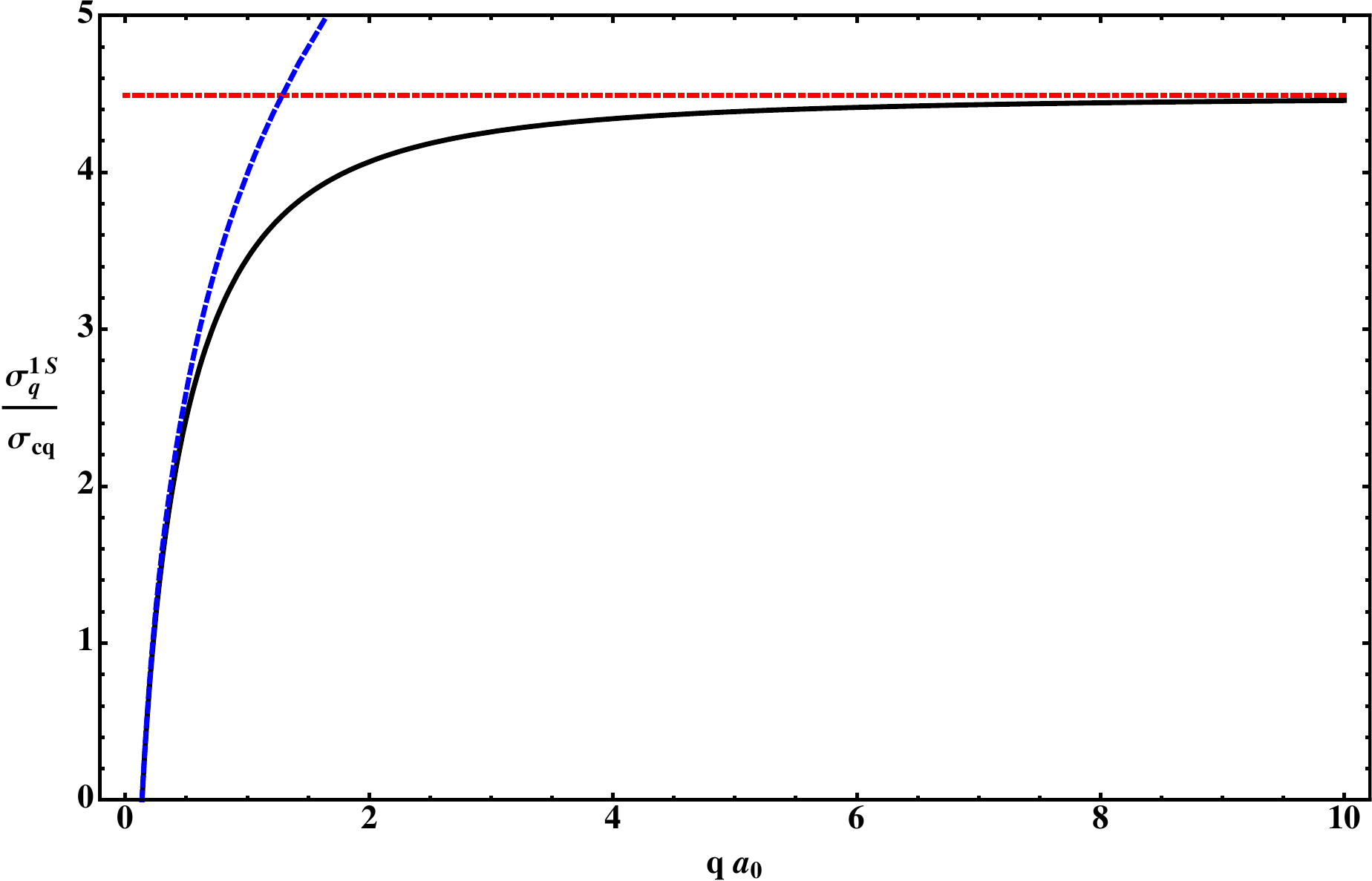}
	\end{center}
	\caption{The cross section for $q+\Psi(1S)\to q+(Q\overline{Q})_8$ \cite{quasifree,miguel}, 
	normalized by the constant $\sigma_{cq}\equiv 32\pi n_f\,\als^2\,a_0^2/3$, where
	 $a_0=3/(2m\als)$ is the Bohr radius, with $m_D a_0=0.1$. The horizontal axis
	is the momentum of the incoming quark in units of $a_0$. The continuous black line
	is the cross section for $T\sim mv$, the dashed blue line is the result for $mv\gg T$, $m_D\gg mv^2$, obtained in the multipole expansion,
	 and the dash-dotted red line is the result for $T\gg mv\gg m_D$. The last two curves
	describe the limiting cases for the first one, which smoothly interpolates between them. We
	recall that typical thermal constituents of the plasma have momenta of order $T$ and hence
	$T\sim mv$ translates into $qa_0\sim 1$, whereas the blue curve is to be trusted for $qa_0\ll 1$ and the red one for $qa_0\gg 1$.}
	\label{fig_plot}
\end{figure}

For what concerns dissociation by inelastic parton scattering, our analysis \cite{quasifree,miguel}
shows that, for processes with a light particle
in the final state,  a factor of $1\pm f_p(q_\mathrm{out})$ should be added to Eq.~\eqref{eq:facw},
where the plus sign applies to the case of the Bose distribution (Bose enhancement)
 an the minus for the Fermi distribution (Pauli blocking).
 Our calculation differs from the one in \cite{hep-ph/0103124,Grandchamp:2002wp}, where the cross section is identified with two times the one for heavy quark-parton scattering  computed in \cite{CERN-TH-2574}, supplemented by momentum-independent thermal masses.
 In our case the Landau damping terms in the EFT approach
encode the square of the diagrams for heavy quark-parton scattering, of those for heavy antiquark-parton scattering and the interference
term, which depends on the properties of the bound state. We show that in the power counting of the weakly-coupled EFT this last term can never be neglected
and hence the use of two times the simple heavy quark-parton scattering cross section is
inappropriate at 
weak coupling. Our cross section furthermore does depend on the temperature in all regimes.
We plot an example in Fig.~\ref{fig_plot} for the process $q+\Psi(1S)\to q+(Q\overline{Q})_8$. The
different curves are the cross section computed in different regimes, which show the validity
of the employed approximations. 

\section{Euclidean quantities}
\label{sec_thermo}
As remarked in the introduction, potential models have often used thermodynamical
free energies measured on the lattice as input. Popular choices have been the 
gauge-invariant
colour-average free-energy, derived from the correlator of two Polyakov loops 
\cite{McLerran:1981pb}, and the colour-singlet free energy \cite{Nadkarni:1986as}, which is gauge dependent.

In \cite{Brambilla:2010xn} the correlator of two Polyakov loops has been analyzed in
detail. A short-distance perturbative calculation was carried out to NNLO. Euclidean pNRQCD
was then employed to show how the correlator can be expressed in terms of
gauge-invariant colour-singlet and colour-octet free energies, which have been computed
at short distances. An important finding is that these free energies are not equal
to the real-time potentials computed in the same temperature regime: not only
are the free energies purely real, in contrast with the complex potentials, but the
real parts of the potentials are themselves different from the free energies.
 The discrepancy has been traced back
to the different boundary conditions in the two cases: for the potentials
one takes a large Minkowskian time to obtain the pole of the bound-state 
propagator, whereas for the free energy one has the Euclidean time spanning
the entire compactified axis.

The effect of periodic boundary conditions on another gauge-invariant quantity
has been analyzed in detail in \cite{Berwein:2012mw} (see also \cite{matthias}).
There, the renormalization of cyclic Wilson loop, i.e. a rectangular Wilson loop spanning
the entire Euclidean time axis, has been studied. This operator had been found
to be UV divergent after charge renormalization in \cite{Burnier:2009bk}. The
analysis of \cite{Berwein:2012mw} shows how this extra divergence is due to 
the periodic boundary conditions, and how its renormalization involves a mixing
with the Polyakov loop correlator. In particular, the difference $W_c-P_c$, $W_c$
being the cyclic loop and $P_c$ the correlator of Polyakov loops, is multiplicatively
renormalizable and could then be another interesting quantity to be measured on the lattice
and compared with perturbative result, in order to test the applicability
of perturbation theory to $Q\overline{Q}$-related quantities.
\section{Non-static and non-perturbative aspects}
\label{sec_aspect}
The calculations reported in sections~\ref{sec_screen} and \ref{sec_pert}
are for a bound state at rest in the medium frame. In \cite{Escobedo:2011ie}
(see also \cite{miguel})
the NR EFT approach was extended to finite relative velocities
between the medium and the (abelian) bound state. All the available
velocity range, from $v\simg 0$ to $v\to 1$, was explored. For muonic
hydrogen, which is the abelian bound state more closely resembling 
quarkonium in a medium, it was found that, in the ultrarelativistic limit and 
for $T> mv$, the imaginary part due to Landau damping loses its
role of main dissociation agent in favour of colour screening, whereas for 
smaller velocities, $0<v<0.9$, the bound state is more easily dissociated 
for increasing velocities.\\
The effect of a finite relative velocity was also explored in lattice
NRQCD in \cite{Aarts:2012ka} (see also \cite{Ryan}) for bottomonium
$S$-wave states. A momentum dependence of the Euclidean correlators
and the reconstructed spectral functions was observed and compared to
the EFT results.

A second topic of interest and focus of recent research is the
non-perturbative determination of the potential. In \cite{Rothkopf:2011db,Burnier:2012az}
the spectral function of the Wilson loop has been determined by
applying the Maximum Entropy Method to lattice data and a complex
potential has then been extracted from it. A different method
has been used in \cite{Bazavov:2012bq,Bazavov:2012fk}, where
a real quantity has been fit from the exponential decay of
correlators of temporal Wilson lines of varying time extent
at a fixed spatial distance. The strong-coupling version
of pRNQCD \cite{Brambilla:1999xf,Brambilla:2004jw} could be generalized
to finite temperatures along the way the weakly-coupled one has; this
would help shed more light on how the proper in-medium potential 
is to be non-perturbatively determined.

\section{Summary and conclusions}
\label{sec_concl}
In this contribution we have briefly illustrated the development 
and a few chosen results of the NR EFT framework for in-medium
quarkonium bound states. From the more theoretical point of view,
we have seen how with EFTs one can give a rigorous QCD definition and derivation
of the potential, bridging the gap with potentials models which appear as leading-order picture
in pNRQCD and its finite-temperature analogues. Furthermore, the power counting
EFTs naturally possess allow one to 
systematically take into account corrections, be it relativistic $1/m^n$ effects ($n\ge 1$), radiative
corrections, etc. and include all medium effects which are relevant for the 
desired accuracy.

For what concerns phenomenology, we have shown how previous results, such as
the complex potential of \cite{Laine:2006ns}, are accommodated in the EFT and
how new results are obtained \cite{Brambilla:2008cx}, such as the estimation of the $\Upsilon(1S)$ 
dissociation temperature \cite{Escobedo:2010tu} given in Sec.~\ref{sec_screen} 
or the calculation 
of the corrections to the potential, the spectrum and the width for a Coulombic
bound state in the regime $mv\gg T\gg mv^2$ \cite{Brambilla:2010vq}, illustrated in Sec.~\ref{sec_pert},
which could be of relevance for the new frontier of $\Upsilon$ phenomenology
in LHC experiments. We have also shown how the thermal width that arises
in the EFT calculations compares with those obtained in phenomenological
approaches based on the convolution of $T=0$ cross sections. For what
concerns gluo-dissociation, the factorization formula holds and the cross
section \cite{Bhanot:1979vb} is a large-$\nc$ limit of the EFT cross section 
\cite{arXiv:1109.5826}. For dissociation by inelastic parton scattering, instead, our factorization
formula differs from the one employed in the literature and a direct comparison
between the cross sections is not possible due to different approximations 
and validity regions \cite{quasifree}. In both cases
the power counting allows to constrain the validity region of these approaches.

The EFT framework  allows also an analysis of the thermodynamical free energies
extracted from correlators of Polyakov loops and widely employed as input for 
potential models. It is found that the colour-singlet and colour-octet free 
energies that can be defined in the Euclidean EFT framework differ from the real
part of the Minkowski-time potentials, the difference being traced back to the different
boundary conditions in the two cases \cite{Brambilla:2010xn}, 
as summarized in Sec.~\ref{sec_thermo}. There it is also shown how periodic 
boundary condition can induce UV divergences in these Euclidean operators, which
then mix under renormalization \cite{Berwein:2012mw}.

Possible extensions of the framework include its generalization to the 
strong-coupling regime, which could help determining which Euclidean
operators, amenable to lattice measurements, can be used to extract, 
possibly through analytical continuation, the real-time potentials
governing the evolution of the bound state.

\acknowledgments
I thank Matthias Berwein, Nora Brambilla, Miguel \'Angel Escobedo, P\'eter Petreczky,
Joan Soto and Antonio Vairo for collaboration. I acknowledge financial support
from the Natural
Science and Engineering Research Council of Canada and from an Institute
of Particle Physics Theory Fellowship.

\end{document}